\documentstyle[11pt,epsf]{article}
\begin{document}
\pagestyle{empty}
\begin{center}
{\Large Meson-induced correlations of nucleons \\
in nuclear Compton 
scattering} \\
\vspace{0.5cm} 
M.-Th.\,H\"utt\footnote{permanent address:
II. Physikalisches Institut der Universit\"at G\"ottingen,\\
\hspace*{0.5cm}
 Bunsenstr. 7-9, 
 D-37073 G\"ottingen, Germany}$^,$\footnote{e-mail: huett@up200.dnet.gwdg.de } 
and A.I.\,Milstein\footnote{e-mail: milstein@inp.nsk.su} \\
 \footnotesize 
       Budker Institute of 
Nuclear Physics, 630090 Novosibirsk, Russia    
       \end{center}
\date{}

\newtheorem{abb}{Figure}
\newcommand{\bi}[1]{\bibitem{#1}}
\newcommand{\fr}[2]{\frac{#1}{#2}}
\newcommand{\fn}{\scriptsize}
\newcommand{\veceps}{\mbox{\boldmath$\epsilon $}}
\newcommand{\vecrho}{\mbox{\boldmath$\rho $}}
\newcommand{\vecxi}{\mbox{\boldmath$\xi $}}
\newcommand{\vdel}{\mbox{\boldmath$\Delta $}}
\newcommand{\unret}[2]{\mbox{$\stackrel{\circ}{#1}{}\!\!^{#2}$}}
\newcommand{\vk}{{\bf k}}
\newcommand{\vr}{{\bf r}}
\newcommand{\vp}{{\bf p}}
\newcommand{\vQ}{{\bf Q}}

\begin{abstract}
The non-resonant (seagull) contribution to the 
nuclear Compton amplitude at low energies
is strongly influenced by nucleon correlations arising from meson exchange.
We study this problem in a modified Fermi gas model, where nuclear 
correlation functions are obtained with the help of perturbation theory.
The dependence of the mesonic seagull amplitude on the nuclear radius is 
investigated and the influence of a realistic nuclear density on this 
amplitude is dicussed.
We found that different form factors appear for the static part
(proportional to the enhancement constant $\kappa $) of the
mesonic seagull amplitude and for the parts, which contain  
the contribution from electromagnetic polarizabilities.
\end{abstract}

\newpage
\pagestyle{plain}
\pagenumbering{arabic}

\section{Introduction}
Nuclear Compton scattering below pion threshold is sensitive to 
low-energy nucleon parameters (e.g. electromagnetic
polarizabilities) as well as nucleon correlations
inside the nucleus. At present no quantitative consistent 
description for all parts of the nuclear Compton 
amplitude below pion threshold exists. 
Therefore, phenomenological models 
have been developed for the different contributions and 
in the last few years several important pieces of information have been 
extracted from the experimental data. At low energies the relative 
strengths of electromagnetic multipoles were analyzed 
\cite{hay,kappa,carbon-lund,nath1} for 
comparison with predictions from multipole sum rules. The interesting 
question, whether the electromagnetic polarizabilities of the nucleon 
inside the nucleus essentially differ from those of the free nucleon, 
has been theoretically addressed \cite{ericson1, ericson3, martin1, hm} 
and experimentally studied
with reasonable accuracy \cite{carbon-lund,nath2,martin3}.

In \cite{martin1} it was suggested to write the total nuclear 
Compton amplitude $T_A$ as a sum of three  
contributions provided by different physical mechanisms 
(see also \cite{molinari,ziegler}):
the collective nuclear excitations (Giant Resonances), 
$R_{GR}(\omega ,\theta )$, the scattering by quasi-deuteron clusters, 
$R_{QD}(\omega ,\theta )$, and the so-called seagull amplitude, 
$S(\omega ,\theta )$,
where $\omega $ is the photon energy and  $\theta $ is 
the scattering angle.
The physical background of this separation is the following: 
Via the optical theorem and a subtracted dispersion relation the scattering 
amplitude in the forward direction is determined up to an 
additive constant by the total photoabsorption cross section. 
This cross section contains resonant structures, which at different 
energies correspond to different excitation mechanisms. 
At low energies 
(up to 30 MeV) absorption is dominated by giant resonances, which can be 
classified due to their electromagnetic multipolarity. At higher 
energies, but still below pion threshold the photon is mainly absorbed by 
two-nucleon clusters, which is known as the quasi-deuteron mechanism.

The seagull amplitude $S$, which has no imaginary part below pion threshold, 
contains contributions from two fundamentally 
different physical sources, the scattering on individual nucleons 
inside the nucleus (${\cal N}(\omega ,\theta )$) 
and the scattering on correlated nucleon pairs 
(${\cal M}(\omega ,\theta )$).
Such correlations occur as a result of the nucleon-nucleon interaction, 
which can be well described in terms of meson exchange between the 
nucleons.
Meson exchange 
leads to specific observable phenomena in Compton scattering.
Best known is the modification of the 
Thomas-Reiche-Kuhn (TRK) sum rule \cite{levinger}, i.e. the appearance 
of the so-called enhancement constant $\kappa $. Also, meson 
exchange currents can imitate a modification of the nucleon 
polarizabilities \cite{friar,hm}. 
This contribution, coming from the polarizabilities of a correlated 
nucleon pair, has to be subtracted in order to 
single out a change of the bound nucleon's polarizabilities from its 
free values. 
The effect of meson exchange currents on the different electromagnetic 
properties of nuclei is thoroughly discussed in e.g. 
\cite{arenh1, brown1, ew, riska}.
A wide variety of model calculations has been carried out for the 
enhancement constant $\kappa $ as well as for the different contributions 
to $\kappa $,, which are theoretically or experimentally accessible 
\cite{brown3,brown4,brownrho}, see also \cite{molinari}. 
Parts of $\kappa $ could be related to parameters of Fermi liquid theory
and to the notion of quasi-particle masses \cite{brownbook}. A compilation
of results can be found in \cite{ew}.
The contributions to $\kappa $ have also been studied in a 
diagrammatic form \cite{ericson1,ericson2}, 
similar to the approach considered here.

At low energies the dependence of the amplitude $\cal M$ 
on momentum transfer 
${\mbox{\boldmath$\Delta $}}=\vk_2 - \vk_1$ is determined
by the distribution of nucleon pairs inside the nucleus.
Here $\vk_1$ and $\vk_2$ are the momenta of the incoming and outgoing 
photons, respectively.
In the case of heavy nuclei 
the scale of nucleon correlations is essentially smaller than the nuclear
radius $R$. Thus, one can expect that the  
${\mbox{\boldmath$\Delta $}}$-dependence of $\cal M$
is similar to the nuclear charge form factor 
$F_1(\Delta )$. 
However, experimental data clearly indicate 
\cite{schelhaas1, pb-lund, kappa} that this 
${\mbox{\boldmath$\Delta $}}$-dependence cannot fully be identified 
with the form factor $F_1$.
In \cite{christillin2, christillin3} it was suggested to use 
for the amplitude $\cal M$
another form factor 
$F_2$ instead of $F_1$.
It has been proposed to apply 
$F_2(\Delta )$=$F_1^2(\Delta /2)$, which corresponds to 
the distribution of uncorrelated
nucleon pairs \cite{riska,nath1}.
A first attempt to quantitatively discuss the function $F_2$ within 
a model calculation has been made in \cite{albmol}.

In \cite{hm} the amplitude $\cal M$ was considered within a modified 
Fermi gas model, in which the deviation of the nucleon wave functions 
from plane waves was taken into account in a perturbative way.
It was shown that $\cal M$ is given by the convolution of a two-body 
spin-isospin correlation function with matrix 
elements corresponding to the amputated irreducible Feynman diagrams for
meson exchange. If the nuclear radius tends to infinity 
the correlation function becomes proportional to the form factor $F_1$.

Among the current experimental data for various nuclei 
contradictions at large angles occur
\cite{nath2,helium-lund,carbon-lund}. As any modification of the 
energy-dependent part of the mesonic seagull amplitude essentially modifies
the angular dependence, a reliable calculation
of $\cal M$ may help to clarify this situation.
Therefore, a thorough discussion of the effect of finite
nuclear size, as well as of a realistic nuclear density is 
highly due.
The present article is devoted to this problem.

\section{Low-energy behaviour of the nuclear Compton amplitude}

Experimentally, the total photoabsorption cross section 
$\sigma _{\gamma A}(\omega )$ below pion threshold can be 
seperated into a giant resonance ($GR$) part and a 
quasi-deuteron ($QD$) part:
\begin{displaymath}
        \sigma _{\gamma A}(\omega )=\sigma _{GR}(\omega )+
        \sigma _{QD}(\omega )\;,\quad \omega <m  ,
\end{displaymath}
where $m$ is the pion mass.
This serves as a means of identifying the resonance parts
of the scattering amplitude
$R_{GR}$ and $R_{QD}$ as
\begin{equation}
        \mbox{Re} \left( {R_{GR}(\omega ,0)-R_{GR}(0,0)} \right)=
        {{\omega ^2} \over {2\pi ^2}}\,P\int\limits_0^\infty  
        {{{\sigma _{GR}(\omega ')\,d\omega '} \over {\omega ^2-\omega '^2}}}
        \label{mesonGR1}
\end{equation}
and 
\begin{equation}
        \mbox{Im} R_{GR}(\omega ,0)={\omega  \over {4\pi }}\,\sigma _{GR}(\omega ) .
        \label{mesonGR2}
\end{equation}
The same is valid for $R_{QD}$.
The seagull amplitude $S$ has an imaginary part only above pion threshold. 

In addition to the fulfillment of a dispersion relation, the main 
constraint on the nuclear Compton amplitude is the low-energy theorem.
It states that the Compton scattering amplitude at 
$\omega =0$ is equal to the (coherent) Thomson limit
\begin{equation}
        T_A(0,\theta )=-\,\veceps _1\cdot \veceps 
        _2\;{{Z^2e^2} 
        \over {AM}}    .
        \label{thomson}
\end{equation}
Here $\veceps _1 $ and  $\veceps _2 $
are the polarisation vectors of the incoming 
and outgoing photons, respectively, $M$ is the nucleon mass and $e$ is the 
proton charge, $e^2 = 1/137$. The quantities $A$ and $Z$ are the nuclear 
mass number and proton number, respectively.
We define the giant resonance part $\kappa _{GR}$ of the enhancement constant 
$\kappa $ via the following relation
\begin{equation}
\frac{1}{2 \pi^{2} }\, \int\limits_{0}^{ \infty } 
\sigma _{GR}(\omega ) d\omega =
 \frac{NZ}{A}
       \frac{e^2}{M} (1+ \kappa _{GR} ).        \label{mesonGR3}
\end{equation} 
Then, taking into account the fact that 
$R_{GR}(\infty ,0)=0$, one obtains 
the low-energy limit of the giant resonance
part of the Compton amplitude from the dispersion relation
(\ref{mesonGR1}):
\begin{equation}
        R_{GR}(0,\theta )=\,\veceps _1\cdot \veceps _2\;{{e^2} 
        \over M}\,\,{ZN \over A}\,(1+\kappa _{GR} )    .
        \label{resonance}
\end{equation}
Similarly, we write 
\begin{equation}
\frac{1}{2 \pi^{2} }\, \int\limits_{0}^{ \infty } 
\sigma _{QD}(\omega ) d\omega =
 \frac{NZ}{A}
       \frac{e^2}{M} \kappa _{QD} .        \label{mesonQD3}
\end{equation} 
for the quasi-deuteron amplitude, which leads to the relation
\begin{equation}
        R_{QD}(0,\theta )=\veceps _1\cdot 
        \veceps _2\;\,{{e^2} \over M}\,{{ZN} \over A}\kappa _{QD}  .
        \label{mesonQD2}
\end{equation}
From (\ref{thomson}), together with (\ref{resonance}) and (\ref{mesonQD2}), 
one obtains 
\begin{equation}
        S(0,\theta )=-\veceps _1\cdot \veceps _2\;{{Ze^2} 
        \over M}\,\left( {1+{N \over A}\kappa } \right)
        \label{mesonS2}
\end{equation}
for the low-energy limit of the seagull amplitude, where 
$\kappa = \kappa _{GR} + \kappa _{QD}$. 
The first term in brackets in eq. (\ref{mesonS2}) is the low-energy limit 
of the so-called kinetic seagull amplitude, which corresponds to the
scattering by individual nucleons. The term in (\ref{mesonS2}) 
proportional to $\kappa $ is the low-energy limit of the mesonic
seagull amplitude $\cal M$. Note that our notations for the seagull 
amplitude differ slightly from \cite{martin1}.

The parameter $\kappa _{GR}$ in eq. (\ref{mesonGR3}) is related 
to the enhancement constant $\unret{\kappa }{}$, which 
appears in the modified Thomas-Reiche-Kuhn (TRK) sum rule:
\begin{equation}
        \int\limits_0^\infty  {\,\unret{\sigma }{E1} (\omega )}\,d\omega =
        2\pi ^2{{e^2} \over M}\,{{NZ} \over A} (1 + \unret{\kappa }{}) ,
        \label{modTRK}
\end{equation}
where $\unret{\sigma }{E1} (\omega )$ is the unretarded
(i.e. obtained in the long-wavelength approximation) electric dipole 
cross section for nuclear photoabsorption.
In order to clarify the relation between 
$\kappa _{GR}$ and $\unret{\kappa }{}$ let us briefly
discuss the effect of retardation.
In an expansion \cite{walecka} of the plane wave for 
the incoming photon into 
terms with definite total angular momentum $l$ and parity
$\lambda =\pm 1$, 
\begin{eqnarray}
         & &\veceps _\lambda \,\exp (i\omega r\cos \theta )= 
        \label{planewave} \\
         & &\sum\limits_{l=1}^\infty  {i^l}\sqrt {2\pi (2l+1)}\left[ 
        {\lambda \,j_l(\omega r)\,{\bf Y}_{ll\lambda }(\theta ,\phi )-
        \mbox{\boldmath$\nabla $}
         \times \left( {j_l(\omega r)\,{\bf Y}_{ll\lambda }
        (\theta ,\phi )} 
        \right)} \right]  , 
        \nonumber
\end{eqnarray}
each term will contain all powers in $\omega ^2$
starting from $\omega ^{2l}$. In eq. (\ref{planewave}) $j_l(\omega r)$ denotes 
the spherical Bessel function and 
${\bf Y}_{ll\lambda }(\theta ,\phi )$ are vector spherical harmonics.
The first term in the brackets on the rhs of (\ref{planewave}) 
corresponds to a magnetic multipole of the photon and the second term to 
an electric one.
When in eq. (\ref{planewave}) the substitution 
$j_l(\omega r)\to (\omega r)^l/(2l+1)!!$ is made, the 
resulting cross section is called ``unretarded''.
If the wavelength of the incoming photon is of the same order as the 
nuclear radius, it is impossible to expand the photon plane wave with 
respect to $\omega r$ in the matrix elements for photoabsorption. At 
these photon energies the effect of retardation is essential.

All absorption cross sections, which in 
principle can be obtained directly from experiment, are by definition
retarded quantities. At high energies the cross sections 
calculated with the use of an unretarded photon wave function differ 
essentially from retarded cross sections and cannot be extracted directly 
from experimental data. Nevertheless, unretarded cross sections are 
convenient objects in theoretical investigations.
If the potential $V$ entering into the Hamiltonian 
contains velocity-dependent or charge exchange contributions, it is well 
known (see e.g. \cite{ew,molinari}) that this 
leads to a modified TRK sum rule (\ref{modTRK}) with 
$\unret{\kappa }{}$ given by
\begin{equation}
        \unret{\kappa }{}={{AM} \over {NZ}}\,\left\langle 0 
        \right|[D_z,[V,D_z]]\left| 0 \right\rangle    .
        \label{kappaunret}
\end{equation}
In (\ref{kappaunret}) $D_z$ is the $z$-component of the intrinsic 
electric dipole operator.
Again, since the unretarded $\unret{\sigma }{E1}$ and 
$\unret{\kappa }{}$ are by themselves not observable, it is necessary to 
establish a connection with experimentally observed quantities. 
This has been attempted by Gerasimov \cite{gerasimov}.
Note that $\unret{\sigma }{E1}$ consists of two parts, a giant resonance 
part $\unret{\sigma }{GDR}$ and a quasi-deuteron part 
$\unret{\sigma }{QD}$. 
For the giant resonance region Gerasimov's argument states that 
in the sum rule (\ref{mesonGR3}) the 
contribution of higher multipoles precisely cancels the 
retardation correction to $\sigma ^{E1}_{GR}$.
Therefore, $\kappa _{GR}$
is equal to the contribution of the giant dipole resonance  
to $\unret{\kappa }{}$. 
A discussion of the applicability of Gerasimov's argument 
at giant resonance energies can be found in \cite{kappa}.

Let us now return to the properties of the seagull amplitude. In addition 
to the static limit (\ref{mesonS2}), it is necessary to take into account 
the corrections proportional to $\omega ^2$. Usually the seagull 
amplitude is represented in the following form 
(see e.g. \cite{nath2,martin1}):
\begin{eqnarray}
        & &S(\omega ,\theta )=-{{Ze^2} \over M}
        \veceps _1\cdot \veceps _2\left( {F_1(\Delta )+
        \kappa {N \over A}F_2(\Delta )} \right)+ \matrix{{}\cr
{}\cr
} 
        \nonumber \\
        & &A\omega ^2\left[ {\bar \alpha _NF_1(\Delta )+
        \delta \alpha F_2(\Delta )} \right]\;\veceps _1\cdot 
        \veceps _2+
        \label{fullseagull} \\
        & &A\left[ {\bar \beta _NF_1(\Delta )+
        \delta \beta \,F_2(\Delta )} \right](\veceps _1\times 
        {\bf k}_1)\cdot (\veceps _2\times {\bf k}_2) \matrix{{}\cr
{}\cr
} ,
        \nonumber
\end{eqnarray}
where $\bar \alpha _N=(Z\bar \alpha _p+N\bar \alpha _n)/A$  is the 
average electric and 
$\bar \beta _N=(Z\bar \beta _p+N\bar \beta _n)/A$ is the average magnetic 
polarizability of the individual nucleon.
The quantities $\delta \alpha $ and $\delta \beta $ are the contribution 
of correlated nucleon pairs to the total electric and magnetic polarizability, 
respectively.
In eq. (\ref{fullseagull}) the $\Delta $-dependence of the seagull 
amplitude is contained in the one-nucleon form factor $F_1$ and the 
two-nucleon form factor $F_2$. The function $F_1$ can be identified with 
the experimentally accessible nuclear charge form factor, while for the 
function $F_2$ phenomenological descriptions have to be made.
In \cite{hm} it was shown that in the limit $R>>1/m$, where again $m$ 
denotes the pion mass,  one has 
$F_2$=$F_1$. Also it was shown that the approximation 
(\ref{fullseagull}), where only terms up to $o(\omega ^2)$ were taken 
into account, reproduces with good accuracy the energy dependence  
up to 100 MeV.
For a finite $mR$ some difference between $F_2$ and $F_1$ 
appears.
Strictly speaking, the form factors for 
$\kappa $, $\delta \alpha $ and $\delta \beta $ in eq. (\ref{fullseagull}) 
also differ from each other.
In the next two sections we study the effect of finite nuclear size on $\kappa 
$, $\delta \alpha $ and $\delta \beta $, as well as the modification of 
form factors with the use of a two-nucleon 
spin-isospin correlation function.
For simplicity we consider symmetric nuclei, $N$=$Z$=$A/2$.

\section{Nucleon correlations and mesonic seagull amplitude}

The mesonic seagull amplitude ${\cal M }$ can be written in the following 
form \cite{ericson1,ericson2,hm}:
\begin{equation}
        {\cal M }=\int {{{d{\bf Q}} \over 
        {(2\pi )^3}}}\;{\cal F}^{ij}({\bf Q})\;T_{ij}({\bf Q}) .
        \label{s2-8}
\end{equation}
The amplitude $T_{ij}$ is determined by amputated diagrams, 
which contain only the nucleon vertices, but not its wave functions.
The diagrams corresponding to the contribution $T^{ij}_{(\pi )}$
of $\pi $-meson exchange to $T_{ij}$ are 
shown in Fig.\,(\ref{fig1}). The explicit form of $T^{ij}_{(\pi )}$
is given in \cite{hm}, eq.\,(8).
It is also necessary to take into account $\rho $-meson exchange. This 
will be done following the prescription of \cite{brownrho}. 
The correlator ${\cal F}^{ij}$ entering into eq.\,(\ref{s2-8}) has the following 
general form:
\begin{equation}
        {\cal F}^{ij}=\sum\limits_{a\ne b} {\left\langle 0 
        \right|\tau _a^{(-)}\tau _b^{(+)}\sigma _a^i\sigma _b^j\;}e^{i\;{\bf Q}\cdot 
        ({\bf x}_b-{\bf x}_a)}\;e^{-i\;{\bf \Delta }
        \cdot ({\bf x}_a+{\bf x}_b)/2}\left| 0 
        \right\rangle   .
        \label{s2-9}
\end{equation}
Here the summation with respect to $a$ and $b$ is performed over all 
nucleons, $\tau _a^{(\pm )} = (\tau _a^1 \pm i \tau _a^2)/2$ are the isospin 
raising and lowering operators, while $\sigma _a ^i/2$ denotes the $i$-th 
component of the spin operator for the $a$-th nucleon.
For the case of a pure Fermi gas model the correlation function is given by
\begin{equation}
        {\cal F}^{ij}_0 = -2\delta ^{ij} \int {{{d{\bf p}_1d{\bf p}_2} 
        \over {(2\pi )^6}}}\int {d{\bf x}_1d{\bf x}_2}\,e^{-i({\bf x}_1+
        {\bf x}_2){\bf \Delta }/2}\,e^{i({\bf x}_1-{\bf x}_2)({\bf p}_1-{\bf 
        Q}-{\bf p}_2)}   .
        \label{a1-1}
\end{equation}
The range of integration for the nucleon momenta ${\bf p}_1$ and
${\bf p}_2$ is the sphere with Fermi momentum $p_F$ as a radius, 
while the integration
with respect to ${\bf x}_1$ and ${\bf x}_2$ is performed 
over the nuclear volume $V$.
Fermi momentum $p_F$ and nuclear volume $V$ are related via
$p_F^3 V$=$3\pi ^2 Z$.
Taking the integral with respect to ${\bf p}_1$ and ${\bf p}_2$ we 
obtain
\begin{equation}
        {\cal F}^{ij}_0 = \delta ^{ij} 
        \int {d{\bf x}_1d{\bf x}_2}\,e^{-i({\bf x}_1+
        {\bf x}_2){\bf \Delta }/2}\,e^{-i({\bf x}_1-{\bf x}_2){\bf Q}}\,\,
        g_C^{(0)}({\bf x}_1-{\bf x}_2) ,
        \label{s0new}
\end{equation}
where 
\begin{equation}
        g_C^{(0)}(\rho )=-2\left( {{{p_F^3} 
        \over {6\pi ^3}}} \right)^2 F^2(p_F\rho )\;,\quad F(x)={3 
        \over {x^2}}\left( {{{\sin x} \over x}-\cos x} \right)  .
        \label{g0f}
\end{equation}
Thus, in a pure Fermi gas model the correlator is proportional to $\delta ^{ij}$,
i.e. only central correlations appear.
However, it is well-known that tensor correlations strongly influence the 
parameters of the mesonic seagull amplitude $\cal M$. For instance, 
tensor correlations give the biggest contribution to the 
value of the enhancement constant $\kappa $ \cite{brown1}.
In order to obtain a quantitative description of the effect of tensor 
correlations, a modified Fermi gas model was considered in \cite{hm}, 
where a correlation function was obtained 
in a perturbative way by evaluating three-nucleon diagrams and 
additional two-nucleon diagrams.
We represent the correlation function, eq.\,(\ref{s2-9}), in the 
following form 
\begin{eqnarray}
         & &{\cal F }^{ij}=\int {d{\bf x}_1d{\bf x}_2}\,e^{-i{\bf \Delta }
         \cdot ({\bf x}_1+{\bf x}_2)/2}e^{-i{\bf Q}\cdot ({\bf x}_1-{\bf x}_2)}
         \times 
        \label{mwithg} \\
        & &\left[ {g_C({\bf x}_1-{\bf x}_2)\;\,\delta ^{ij}+
        g_T({\bf x}_1-{\bf x}_2)\;t^{ij}} \right] ,
        \nonumber
\end{eqnarray}
where 
\begin{equation}
        t^{ij}={{3Q^iQ^j} \over {Q^2}}-\delta ^{ij}  .
        \label{s2-11}
\end{equation}
In eq.(\ref{mwithg}) the functions $g_C$ and $g_T$ describe the 
central and tensor correlations of two 
nucleons, while the exponential function depending on ${\bf \Delta }$
is responsible for the distribution of such nucleon pairs inside the 
nucleus.
The functions $g_C$ and $g_T$ are related to momentum space 
correlation functions
${\cal F }_C$ and ${\cal F }_T$ obtained in \cite{hm} via
\begin{equation}
        g_{C,T}(\rho )= \frac{1}{V} \int {{{d{\bf Q}} 
        \over {(2\pi )^3}}}\,{\cal F}_{C,T}(Q)\,e^{i\,
        \vecrho \cdot {\bf Q}}  .
        \label{gct}
\end{equation}
Note that in the calculation of ${\cal F }_C$ and ${\cal F }_T$ 
the contribution from $\rho $-meson exchange has been taken 
into account.
Next, we expand $T^{ij}_{(\pi )}$ in eq. (\ref{s2-8}) 
with respect to ${\bf k}_1$ and ${\bf k}_2$ up to
$o(\omega ^2)$, pass to variables $\vecrho = {\bf x}_2 - {\bf 
x}_1$ and  $\vecxi = ({\bf x}_1 + {\bf x}_2)/2$.
Then, taking the integral with respect to ${\bf Q}$ and the angles
of $\vecrho $ and $\vecxi $ we obtain 
\begin{eqnarray}
        {\cal M }&=&{{Ae^2} \over {4M}}\,\left\{ {\matrix{{}\cr
{}\cr
}} \right.\Phi _1(\Delta )\,\veceps _1\cdot \veceps _2+
{{\omega ^2} \over {m^2}}\,\Phi _2(\Delta )\,\veceps _1\cdot 
\veceps _2+
        \nonumber \\
        & &{1 \over {m^2}}\Phi _3(\Delta )\,(\veceps _1\times 
        {\bf k}_1)\cdot (\veceps _2\times {\bf k}_2)\left. {\matrix{{}\cr
{}\cr
}} \right\}  ,
        \label{mphi}
\end{eqnarray}
where
\begin{eqnarray}
        & &\Phi _i={{2Mf^2} \over {3m\pi ^2}}(2Rp_F)^3
        \int\limits_0^1 {dx}\,\left[ {G_i^C(\rho _1)\,\tilde g_C(\rho _2)+} 
        \right.\left. {G_i^T(\rho _1)\,\tilde g_T(\rho _2)} \right]\times 
        \nonumber \\
        & &x^2 e^{-\rho _1}\left( {\int\limits_0^{1-x} {d\xi }\;\xi \,{{\sin \xi R\Delta } 
        \over {R\Delta }}+\int\limits_{1-x}^{\sqrt {1-x^2}} 
        {d\xi }\;{{\sin \xi R\Delta } \over {R\Delta }}\;{{1-x^2-\xi ^2} 
        \over {2x}}} \right)  ,
        \label{phi}
\end{eqnarray}
with $f$ being the pion-nucleon coupling constant, $f^2/4\pi$=0.08.
In eq. (\ref{phi}) the following abbreviations have been used:
\begin{displaymath}
        \tilde g_{C,T}(\rho _2 )={1 \over 2}\left( {{{6\pi ^2} 
        \over {p_F^3}}} \right)^2
    g_{C,T}(\rho _2/p_F )\;,\quad \rho _1=2Rmx\;,\quad 
        \rho _2=2Rp_Fx\quad .
\end{displaymath}
The functions $G_i^{C,T}$ are of the form
\begin{eqnarray}
         &  & G_1^C(\rho _1 )=\rho _1 \;,\quad G_1^T(\rho _1 )=
         {{2\rho _1 ^2-12} \over \rho _1 }\;,
        \nonumber \\
         &  & G_2^C(\rho _1 )={{150+30\rho _1 -\rho _1 ^3} \over {60}}\;,
         \quad G_2^T(\rho _1 )=
         {{24+12\rho _1 -\rho _1 ^3} \over {30}}
        \nonumber \\
         &  & G_3^C(\rho _1 )={{3-21\rho _1 +2\rho _1 ^2} 
         \over {12}}\;,\quad G_3^T(\rho _1 )={{2\rho _1 ^2-3\rho _1 -15} \over 6}
        \nonumber
\end{eqnarray}
The integral with respect to $\xi $ 
in (\ref{phi}) can easily be taken analytically, but we represent the
result in this form for the sake of brevity. 
By comparing eq. (\ref{mphi}) with the corresponding terms in 
eq. (\ref{fullseagull}) one sees that 
the parameters appearing in the mesonic seagull amplitude
are given by the functions 
$\Phi _i(\Delta )$ at $\Delta $=0:
\begin{equation}
        \kappa =-\Phi _1(0)\;,\quad \delta \alpha ={{e^2} 
        \over {4Mm^2}}\Phi _2(0)\;,\quad \delta \beta ={{e^2} 
        \over {4Mm^2}}\Phi _3(0)  .
        \label{phinorm}
\end{equation}
It is evident from eq. (\ref{phi}) that three different form factors
\begin{equation}
        F_2^{(i)}(\Delta )={{\Phi _i(\Delta )} 
        \over {\Phi _i(0)}} 
        \label{formphi}
\end{equation}
appear instead of only $F_2$.
In the case of magnetic polarizability $\delta \beta $ the contribution 
of the $\Delta $-isobar excitation to the mesonic seagull amplitude 
should also be taken into account 
\cite{arenh2,hm}. 
In our notation this corresponds to an additional contribution 
to the function $\Phi _3$, which is of the following form:
\begin{eqnarray}
        & &\delta \Phi _3={{8Mf_\Delta f_{\gamma N\Delta}f(2Rp_F)^3} 
        \over {81(M_\Delta -M)\pi ^2}}
        \int\limits_0^1 {dx}\,\left[ {G_\Delta ^C(\rho _1)\,\tilde g_C(\rho _2)+} 
        \right.\left. {G_\Delta ^T(\rho _1)\,\tilde g_T(\rho _2)} \right]\times 
        \nonumber \\
        & &x^2 e^{-\rho _1}\left( {\int\limits_0^{1-x} 
        {d\xi }\;\xi \,{{\sin \xi R\Delta } 
        \over {R\Delta }}+\int\limits_{1-x}^{\sqrt {1-x^2}} 
        {d\xi }\;{{\sin \xi R\Delta } \over {R\Delta }}\;{{1-x^2-\xi ^2} 
        \over {2x}}} \right)  ,
        \label{phidelta}
\end{eqnarray}
where $M_\Delta $ is the $\Delta $-isobar mass and
\begin{displaymath}
     G_\Delta ^C(\rho _1)={{6-12\rho _1} 
        \over {\rho _1}}\;,\quad G_\Delta ^T(\rho _1)={{12-6\rho _1} 
        \over {\rho _1}} .
\end{displaymath}
The coupling constants appearing in eq. (\ref{phidelta}) are taken to
be $f_\Delta $=$2f$ and $f_{\gamma N\Delta}$=0.35.

Up to now we have considered a constant 
nucleon density $n_0=p_F^3/3\pi ^2$ 
inside the nucleus, which is normalized as $n_0 V$=$Z$.
Using a local-density approximation we will extend our 
consideration to realistic nuclear densities $n(r)$.
With the help of the usual plane-wave expansion via Legendre polynomials 
$P_l(x)$ we obtain
\begin{eqnarray}
        & &\Phi _i^{(rd)}={{64\pi Mf^2} 
        \over {3mZ}}\,\int\limits_0^\infty  {dx}\,x^2e^{-2mx}\int\limits_x^\infty  
        {dr}\,r^2n^2(r)\,\;\times 
        \label{phireal} \\
        & &\left[ {G_i^C(2xm)\,\tilde g_C(2xp(r))+
        G_i^T(2xm)\,\tilde g_T(2xp(r))} \right]\;\times 
        \nonumber \\
        & &\sum\limits_{l=0}^\infty  {j_l(r\Delta )\,j_l(x\Delta )
        \left( {P_{l-1}(x/r)-P_{l+1}(x/r)} \right)}  ,
        \nonumber
\end{eqnarray}
where $p(r)$=$(3\pi ^2 n(r))^{1/3}$ is the local Fermi momentum.

Eqs. (\ref{mphi})-(\ref{phireal}) form 
the starting point of our numerical investigation of the mesonic seagull 
amplitude.

\section{Numerical results}

The dependence of the normalized correlation functions $\tilde g_C$ and 
$\tilde g_T$ on distance in units of $1/p_F$ is shown
in Fig. (\ref{fig2}). For comparison, the pure Fermi gas prediction, eq. 
(\ref{g0f}), for $\tilde g_C$ is also shown. At distances below 
0.8 fm the difference between this zero-order approximation and the 
result $\tilde g_C$ of our model becomes most significant.
For finite-size (as opposed to point-like) nucleons one may expect
that any correlation function vanishes at distance equal to zero.
However, we checked numerically that 
the absence of this behaviour in our model has only a very
small influence (less than eight per cent) 
on the explicit values for $\kappa $, $\delta \alpha $ and
$\delta \beta $.
Note that due to the use of either the Fermi gas model or a local 
density approximation the accuracy of our results decreases with
decreasing $Z$.

In order to account for the contribution of $\rho $-meson exchange 
to $T^{ij}$ we follow the prescription of \cite{brownrho} and 
make the substitution 
$f^2\rightarrow 2\tilde f^2_\rho$ for the central part of each 
quantity and $f^2\rightarrow -\tilde f^2_\rho$ for the tensor 
part. The pion mass $m$ is substituted in all cases by the 
$\rho $-meson mass $m_\rho $. We used  
$\tilde f^2_\rho $=$0.4 f^2_\rho $ 
and $f^2_\rho /4\pi = 4.86 $. 
In accordance with \cite{brownrho,brown2} the coefficient 0.4 
approximates the influence of short-range repulsive correlations 
due to the exchange of $\omega $- and $\sigma $-mesons.
  
We now pass to the discussion of 
our numerical results for the different contributions to 
$\kappa $. Note that, as it was argued in \cite{brown3} and is also 
discussed in \cite{ew}, the main contribution to $\kappa ^{GR}$ comes from 
central correlations, while  $\kappa ^{QD}$ is mainly determined by 
tensor correlations.
In the pure Fermi gas model, where $g_C$= $g_C^{(0)}$ and 
$g_T$=0, the contribution to $\kappa $ from pion exchange $\kappa ^\pi$
is approximately equal to the $\rho $-meson contribution $\kappa ^\rho $.
For nuclear matter (infinite nuclear radius) one has 
$\kappa ^\pi$=$\kappa ^\rho$=0.2, which is in agreement with a variety of
model calculations, e.g. \cite{brownrho, brown3, brown4}. 
For finite nuclei, again 
in the pure Fermi gas, the value for  $\kappa ^\pi$ decreases slightly 
with decreasing $Z$, whereas $\kappa ^\rho$ remains the same.
With inclusion of the full correlation functions $g_C$ and $g_T$ from 
Fig. (\ref{fig2}) the situation for $\kappa $ changes drastically. 
Now the main contribution to $\kappa $ comes from tensor correlations 
related with pion exchange. The pionic central contribution is still 
of the same order as before, while $\kappa _C^\rho $ becomes 
negligible. The only significant contribution from $\rho $-meson 
exchange is now due to tensor correlations and has a negative value.
All these relations between the different ingredients to $\kappa $
remain valid, when realistic nuclear densities are considered.
In Fig. (\ref{fig3}) the different contributions to $\kappa $ are shown
as a function of $Z$ for the modified Fermi gas model. 
The realistic-density result $\kappa ^{(rd)}$, which is obtained from 
eq. (\ref{phireal}), is also shown in
Fig. (\ref{fig3}). For the densities $n(r)$ 
we used a three-parameter Fermi parametrization with values 
for the different nuclei taken from \cite{datatab}.

In the case of electric and magnetic polarizabilities 
$\delta \alpha $ and $\delta \beta $ the contributions from 
$\rho $-meson exchange are suppressed by a factor of $m^2/m_\rho ^2$
in comparision with the pion contributions and, therefore, are negligible.
The values of $\delta \alpha $ and $\delta \beta $ are determined 
mainly by pionic central correlations, as can be seen in Figs. (\ref{fig4})
and (\ref{fig5}). In the case of $\delta \beta $ the inclusion of
the $\Delta $-isobar intermediate state produces a noticable effect
(cf. Fig. (\ref{fig5})). 
In this contribution the values due to central and tensor 
correlations are of the same order.
Note that $\kappa $, $\delta \alpha $ and
$\delta \beta $ get close to their asymptotic (nuclear matter) 
values calculated in \cite{hm} only at extremely high $Z$. 
The size of both, $\delta \alpha $ and $\delta \beta $ becomes
noticably smaller, when a realistic density is taken into account. 
This effect gains importance with decreasing $Z$. 
The ratio of central and tensor contributions to $\delta \alpha $
is approximately the same for a realistic density as in a modified
Fermi gas model. For $\delta \beta $ the influence of tensor
correlations in the realistic-density case is slightly stronger than
for homogeneous nuclear density.

We consider now the dependence of the mesonic seagull amplitude
$\cal M$ on momentum transfer $\Delta $, which is determined 
by the form factors $F_2^{(i)}(\Delta )$ (cf. eq. (\ref{formphi})).
In order to cover a wide range of $Z$, results for the form factors will 
be given for lead ($A/2$=104), calcium ($A/2$=20) and carbon ($A/2$=6).
Our results indicate that $F_2^{(2)}$ for the term proportional to 
$\delta \alpha $ and $F_2^{(3)}$ (for $\delta \beta $) are 
equal with high accuracy, but differ significantly from 
the form factor $F_2^{(1)}$ for the 
term containing $\kappa $.
All three functions  $F_2^{(i)}$ differ noticably from $F_1$. 
Figs. (\ref{fig6}),(\ref{fig7}) and (\ref{fig8}) show the 
corresponding curves for lead, calcium and carbon, respectively,
for the case of a realistic density.
One can see that the frequently used phenomenological approximation 
$F_2(\Delta )$=$F_1^2(\Delta /2)$, which is also shown in 
Figs. (\ref{fig6})-(\ref{fig8}), is not in agreement 
with the $\Delta $-dependence 
of the amplitude $\cal M$ obtained here.
For very small $\Delta $ it is convenient to represent the form 
factors as $F_2^{(i)} = 1 - \Delta ^2 r^2_i/6$.
Then, for the case of carbon we have $r_1$=1.9\,fm, $r_2$=1.4\,fm.
For calcium we find $r_1$=3.0\,fm and $r_2$=2.5\,fm. For lead
the corresponding values are $r_1$=5.0\,fm and $r_2$=4.7\,fm.
In the case of lead, the result from \cite{albmol} coincides 
within good accuracy with the form factor $F_2^{(2)}$ shown in 
Fig. (\ref{fig6}).

In Fig. (\ref{fig9}) the form factors obtained 
for a homogeneous nuclear density are compared with
those for a realistic density in the case of calcium.
 
\section{Conclusion}
In the frame of our model we demonstrated that central and 
tensor correlations have a strong influence on 
the parameters appearing in the mesonic seagull amplitude.
Our calculation is based on correlation functions obtained with
the help of perturbation theory. Therefore, our predictions 
may still be influenced by higher-order effects in the 
correlation functions. However, we suppose that our model 
describes correctly the role of mesonic effects 
in low-energy nuclear Compton scattering.

The values of the parameters $\kappa $, $\delta \alpha $ and
$\delta \beta $ for finite nuclear size differ
essentially from those obtained for infinite nuclear matter. 
Our calculation indicates the necessity of applying two 
different exchange form factors.
While $F_2^{(1)}$ enters $\cal M$ at the term proportional to $\kappa $,
the form factor $F_2^{(2)}$ is related with the terms containing
the electromagnetic polarizability modifications 
$\delta \alpha $ and $\delta \beta $.

\section*{Acknowledgments}
We are grateful to A.I. L'vov and M. Schumacher for useful discussions. 
We are also indebted to A.M. Nathan for his interest in our work.
M.T.H. wishes to thank the Budker Institute of Nuclear Physics, 
Novosibirsk, for the kind hospitality accorded him during his stay, 
when this work was done.
This work was supported by Deutsche Forschungsgemeinschaft, 
contract 438/113/173.

\newpage
\section*{Figure Captions} 
        \begin{abb}
        \protect\label{fig1}
            Typical diagrams contributing to $T^{ij}_{(\pi )}$. The wavy 
            lines denote photons and dashed lines denote pions. The 
            amputation indicates that $T^{ij}$ contains only the nucleon vertices, but 
            not its wave functions.
        \end{abb}

        \begin{abb}
        \protect\label{fig2}
            Normalized correlation functions $\tilde g_C$ (dashed curve) and
            $\tilde g_T$ (full curve) as a function of $y=\rho p_F$. For 
            comparison the normalized correlation function for a pure 
            Fermi gas model (cf. eq. (\ref{g0f})) is shown (dotted curve).
        \end{abb}

        \begin{abb}
        \protect\label{fig3}
            Dependence of enhancement constant $\kappa $ on proton number $Z$.
            The dashed curve corresponds to the pionic tensor contribution 
            $\kappa _T^\pi $, the dotted curve includes also
            the central contribution $\kappa _C^\pi $ and the dash-dotted
            curve gives the total $\kappa $, including the 
            contribution from $\rho $-meson exchange. 
            The realistic-density result $\kappa ^{(rd)}$ for the full 
            enhancement constant  (cf. eq. (\ref{phireal})) 
            is shown as a full curve.
        \end{abb}

        \begin{abb}
        \protect\label{fig4}
            Pion-exchange contribution to electric polarizability 
            $\delta \alpha $ as a function of $Z$.
            The dashed curve corresponds to the central contribution 
            $\delta \alpha _C$ and the dotted
            curve gives the total $\delta \alpha  
            $=$\delta \alpha _C$+$\delta \alpha _T$.
            The use of a realistic density leads to the full curve.
        \end{abb}

        \begin{abb}
        \protect\label{fig5}
            Pion-exchange contribution to magnetic polarizability 
            $\delta \beta $ as a function of $Z$.
            The dashed curve corresponds to the central contribution 
            $\delta \beta _C$ and the dotted
            curve gives the sum $\delta \beta _C$+$\delta \beta _T$.
            Adding the contribution of the $\Delta $-isobar excitation
            as given in eq. (\ref{phidelta}) leads to the total
            value of $\delta \beta $ given as the dash-dotted curve.
            The use of a realistic density leads to the full curve.
        \end{abb}

        \begin{abb}
        \protect\label{fig6}
            Form factors $F_2^{(i)}(\Delta )$ for $A/2$=104. The dashed 
            curve is $F_2^{(1)}$ and the full curve is $F_2^{(2)}$. For
            comparison the (experimental) charge form factor $F_1$ is
            also shown (dash-dotted curve), as well as the 
            function $F_1^2(\Delta /2)$ (dotted curve).
        \end{abb}

        \begin{abb}
        \protect\label{fig7}
            Same as Fig. (\ref{fig6}), but for $A/2$=20.
        \end{abb}

        \begin{abb}
        \protect\label{fig8}
            Same as Fig. (\ref{fig6}), but for $A/2$=6.
        \end{abb}

        \begin{abb}
        \protect\label{fig9}
            Comparison of form factors $F_2^{(i)}(\Delta )$
            for the realistic-density approach with those 
            for a homogeneous nuclear density. The dashed (dotted)
            curve is $F_2^{(1)}$ for realistic (homogeneous) density,
            while the dash-dotted (full) curve corresponds to 
            $F_2^{(2)}$ for realistic (homogeneous) density.
        \end{abb}

\newpage

\begin{figure}[t]
    \caption{}
    \vspace{2cm}
    \epsfxsize=16cm
    \centerline{\epsfbox{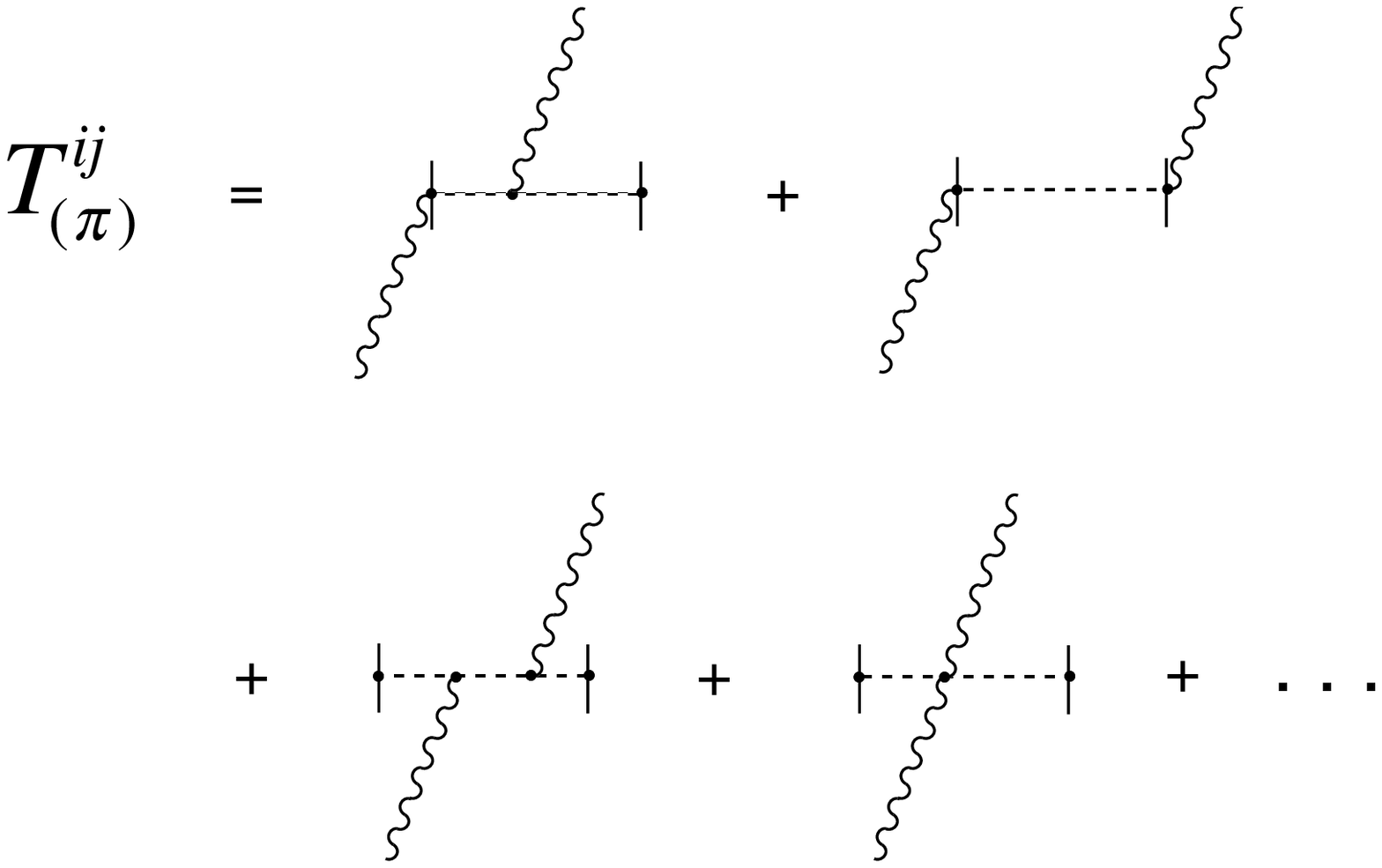}}
\end{figure}

\newpage

\begin{figure}[t]
    \caption{}
    \vspace{2cm}
    \epsfxsize=16cm
    \centerline{\epsfbox{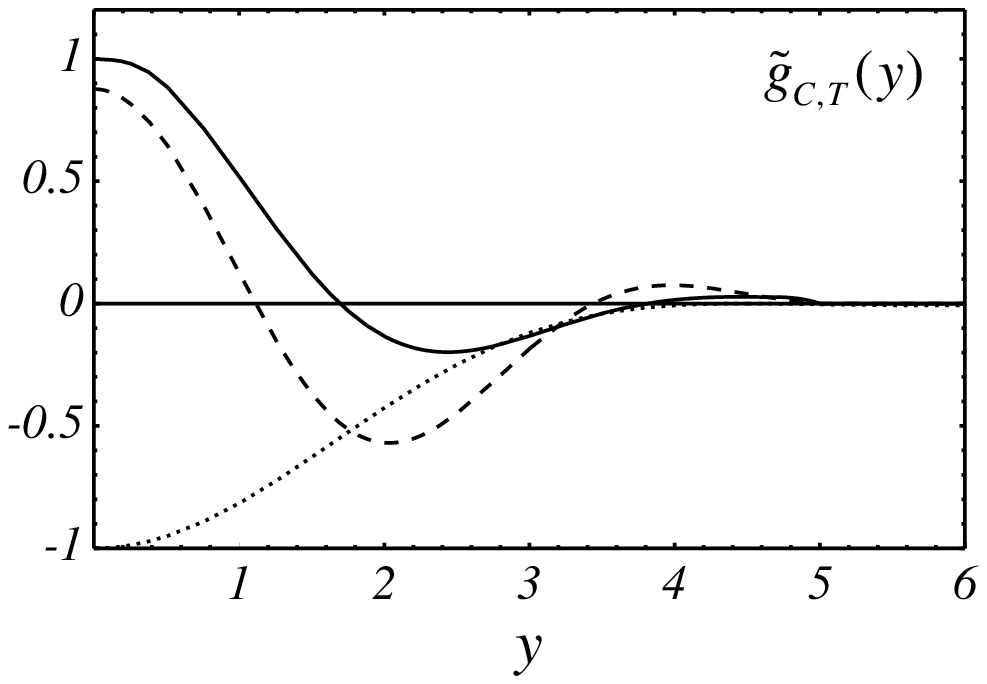}}
\end{figure}

\newpage

\begin{figure}[t]
    \caption{}
    \vspace{2cm}
    \epsfxsize=16cm
    \centerline{\epsfbox{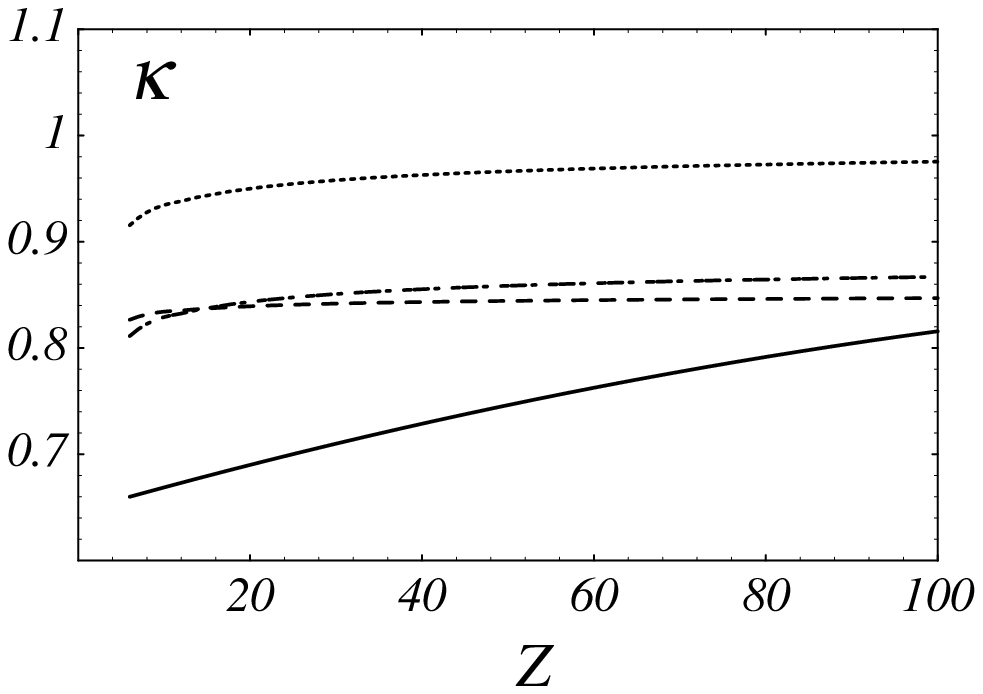}}
\end{figure}

\newpage

\begin{figure}[t]
    \caption{}
    \vspace{2cm}
    \epsfxsize=16cm
    \centerline{\epsfbox{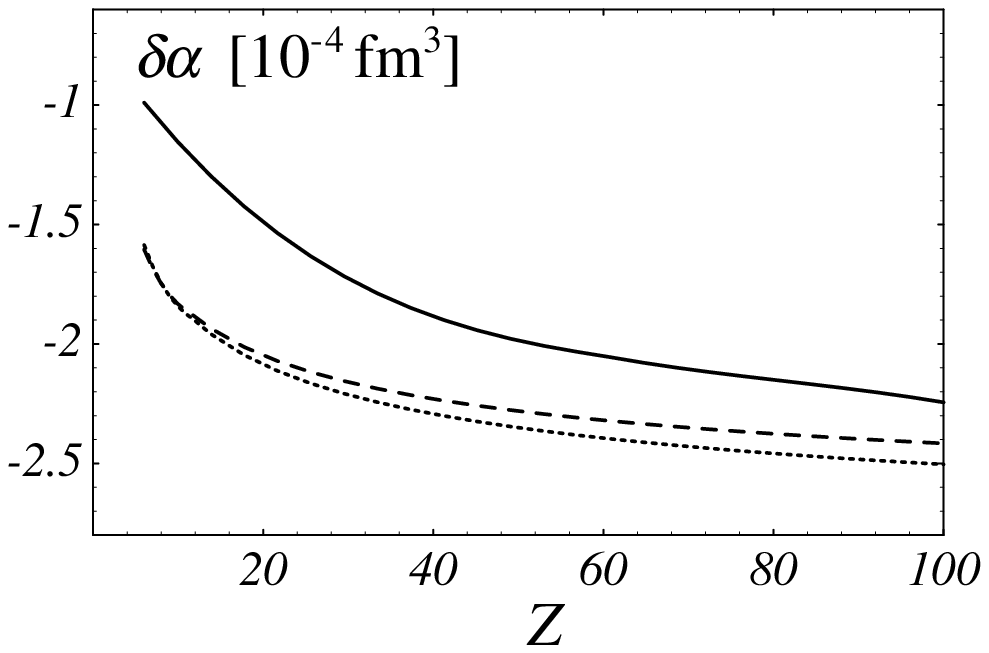}}
\end{figure}

\newpage

\begin{figure}[t]
    \caption{}
    \vspace{2cm}
    \epsfxsize=16cm
    \centerline{\epsfbox{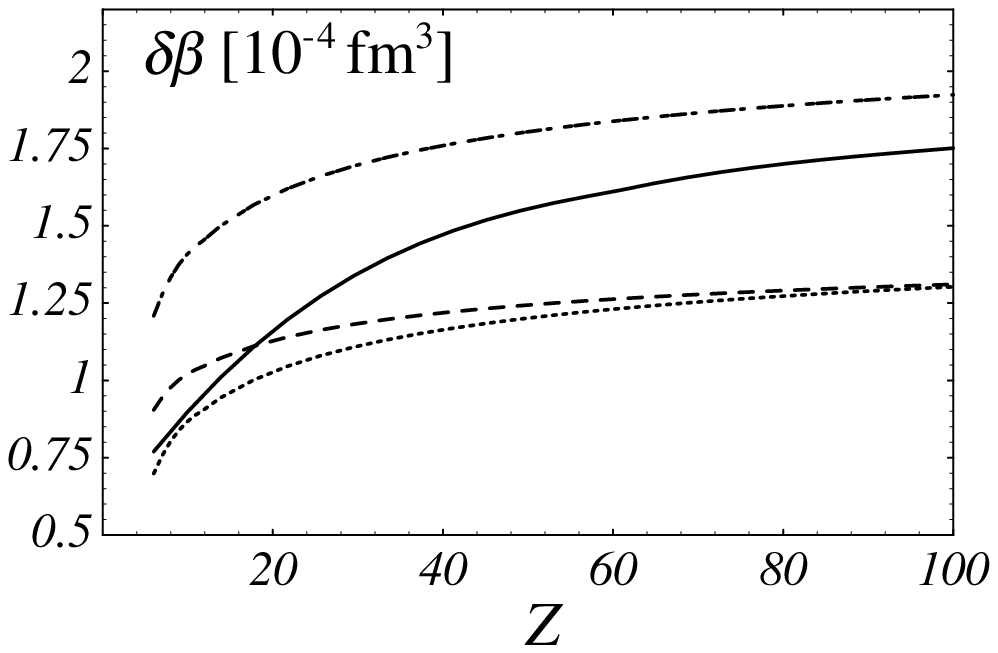}}
\end{figure}

\newpage

\begin{figure}[t]
    \caption{}
    \vspace{2cm}
    \epsfxsize=16cm
    \centerline{\epsfbox{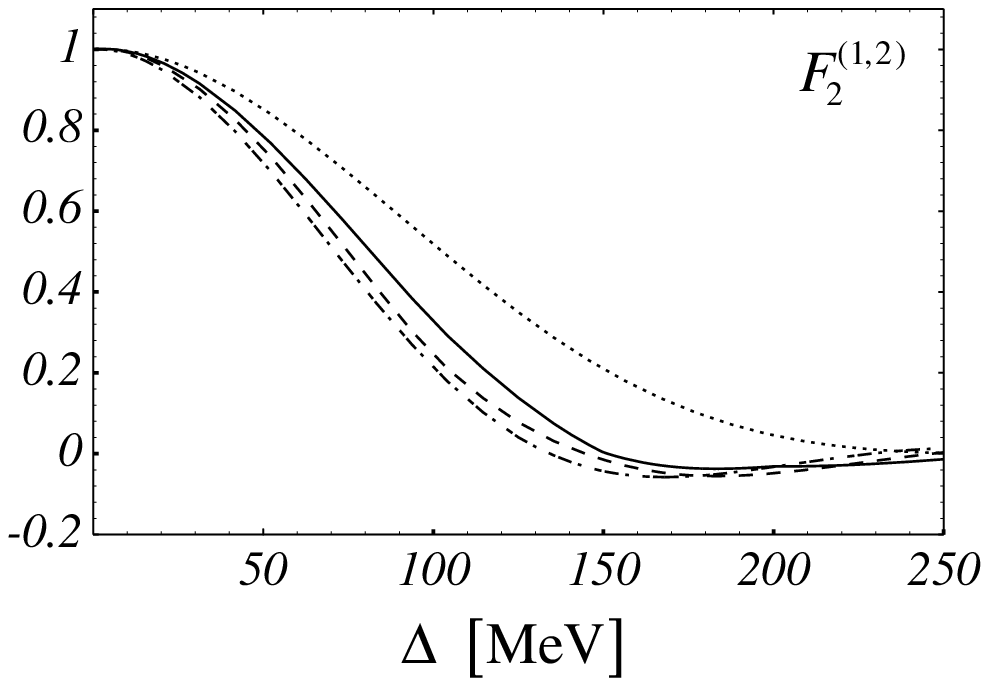}}
\end{figure}

\newpage

\begin{figure}[t]
    \caption{}
    \vspace{2cm}
    \epsfxsize=16cm
    \centerline{\epsfbox{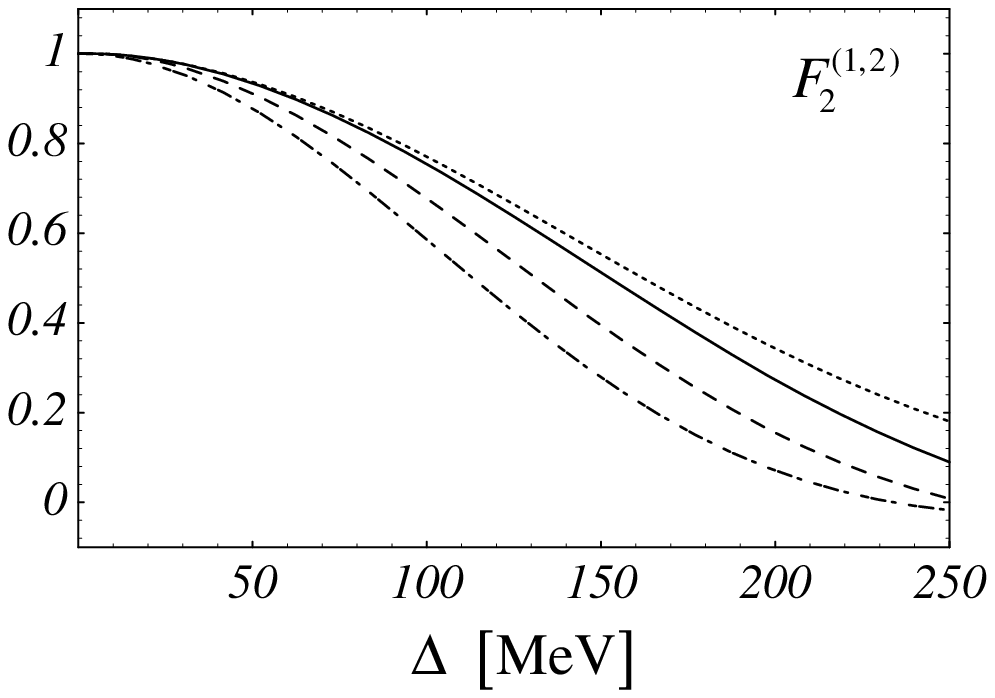}}
\end{figure}

\newpage

\begin{figure}[t]
    \caption{}
    \vspace{2cm}
    \epsfxsize=16cm
    \centerline{\epsfbox{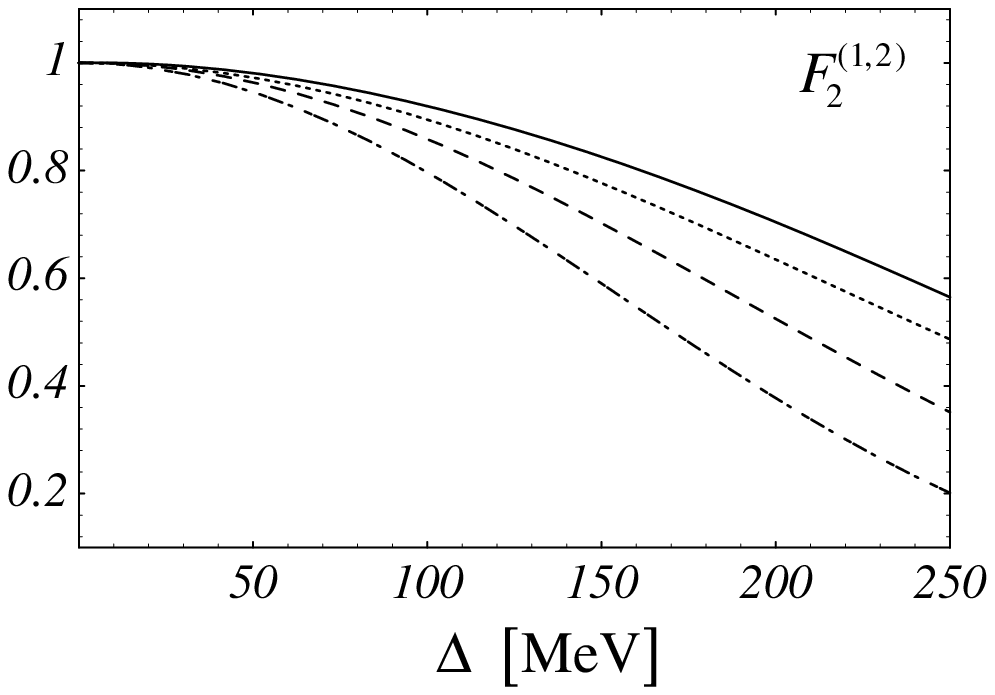}}
\end{figure}

\newpage

\begin{figure}[t]
    \caption{}
    \vspace{2cm}
    \epsfxsize=16cm
    \centerline{\epsfbox{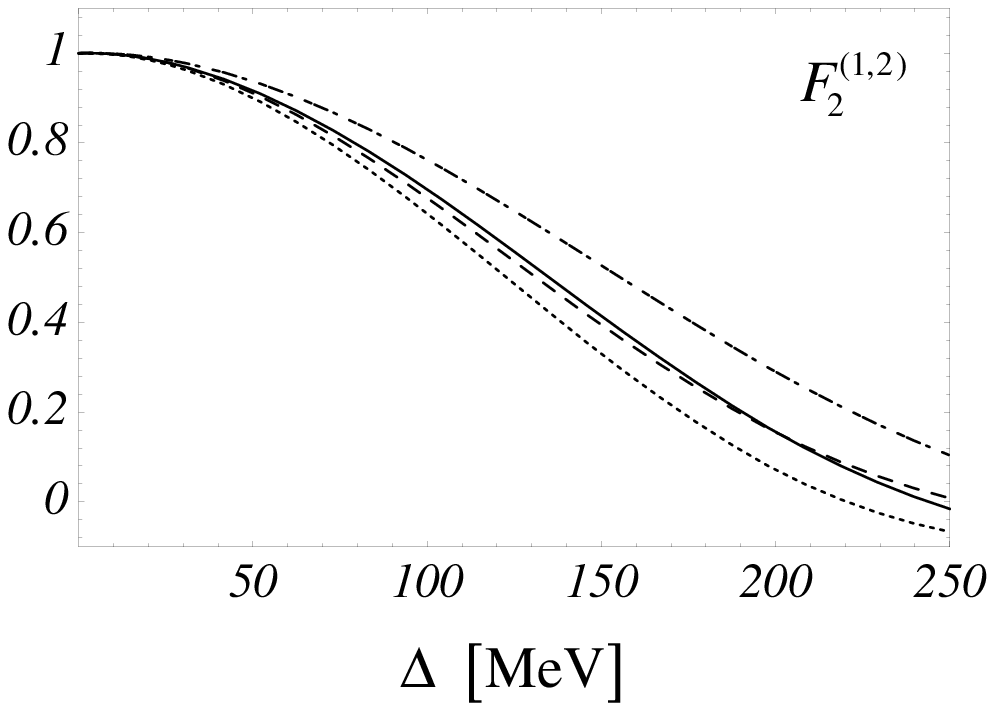}}
\end{figure}
\end{document}